# A New Theory of Cosmology That Preserves the Generally Recognized Symmetries of Cosmos, Explains the Origin of the Energy for Matter Field, but Excludes the Existence of the Big Bang

**Fang-Pei Chen**
Department of Physics, Dalian University of Technology, Dalian 116024, China.
E-mail: chenfap@dlut.edu.cn

**Abstract**

While the generally recognized symmetries of cosmos are preserved, conservation laws for gravitational system are reconsidered and the Lagrangian density of pure gravitational field is revised. From these considerations, some of the theoretical foundations of the current cosmology are extended or revised, and a new theory of cosmology is established. This new theory leads to the following distinct properties of cosmos: the energy of matter field might originate from the gravitational field; the big bang might not have occurred; the fields of the dark energy and some parts of the dark matter would not be matter fields but might be gravitational fields, they would only interact with gravitational force but could not interact with other forces. These distinct properties can be tested by future experiments and observations.

## 1. Introduction

The main theoretical foundations of the current cosmology may well be summarized into the following three statements [1]:

**Statement (1).** The gravitational field variable is $g_{\mu\nu}(x)$, the Lagrangian density of pure gravitational field is $\sqrt{-g}\, L_G(x) = -\dfrac{\sqrt{-g}}{16\pi G} R(x)$, from which the Einstein's field equations

$$R^{\mu\nu} - \frac{1}{2} g^{\mu\nu} R = -8\pi G\, T^{\mu\nu}_{(M)} \qquad (1)$$

can be derived; where $T^{\mu\nu}_{(M)} \overset{def}{=} \dfrac{2}{\sqrt{-g}} \dfrac{\delta(\sqrt{-g}\, L_M)}{\delta g_{\mu\nu}}$ is the energy-momentum tensor for matter field[2], $\sqrt{-g}\, L_M(x)$ is the Lagrangian density of matter field and it includes the interaction between matter field and gravitational field [3], so that $T^{\mu\nu}_{(M)}$ includes the interaction energy between matter field and gravitational field. The action integral for matter field $I_M = \int \sqrt{-g}\, L_M d^4 x$ and the action integral for gravitational field

$$I_G = \int \sqrt{-g}\, L_G d^4 x$$ are all invariant under the local nonhomogeneous Lorentz group [3].

In addition, $I_G$ is invariant only under space-time symmetry, but $I_M$ is invariant under both space-time symmetry and internal symmetry, so the gravitational field can be acted only by gravitational force and can not be acted by other forces, but the matter field can be acted by both gravitational force and other forces [6].

**Statement (2)**. The universe is assumed to be spatially homogeneous and isotropic; this assumption is called cosmological principle. It means that the three-dimensional space is maximally symmetric subspace of the whole of space-time [1], so the universe has the Robertson-Walker metric

$$d\tau^2 = -dt^2 + a(t)^2 \left\{ \frac{dr^2}{1-kr^2} + r^2 d\theta^2 + r^2 \sin^2\theta d\phi^2 \right\} \qquad (2)$$

where $(r,\theta,\phi)$ are comoving space coordinates and $a(t)$ is the scale factor. The constant $k =$ -1, 0, 1, is used to indicate the spatial curvature. The homogeneous and isotropic assumption also implies that $T^{\mu\nu}_{(M)}$, the energy-momentum tensor of the matter field, should take the form of ideal fluid[1]:

$$T^{\mu\nu}_{(M)} = (\rho_M + p_M) u^\mu u^\nu + p_M g^{\mu\nu} \qquad (3)$$

where $u^\mu$ is the 4-velocity of matter, $\rho_M$ is the mean energy density of matter, and $p_M$ is the mean pressure of matter.

**Statement (3)**. The contracted Bianchi identities [1]

$$(R^{\mu\nu} - \frac{1}{2} g^{\mu\nu} R)_{;\mu} = 0 \qquad (4)$$

are obeyed in all cases; Eq. (1) and Eq. (4) then give rise to $T^{\mu\nu}_{(M);\mu} = 0$, these relations can be transformed into [4]

$$T^\mu_{(M)\nu;\mu} = \frac{1}{\sqrt{-g}} \frac{\partial(\sqrt{-g}\, T^\mu_{(M)\nu})}{\partial x^\mu} - \frac{1}{2} \frac{\partial g_{\mu\rho}}{\partial x^\nu} T^{\mu\rho}_{(M)} = 0 \quad . \qquad (5)$$

On the other hand, because $I_G = \int \sqrt{-g}\, L_G d^4 x$ is invariant under the local space-time translation group, there are the other identities [5, 10]

$$\frac{\sqrt{-g}}{16\pi G}(R^{\mu\nu}-\frac{1}{2}g^{\mu\nu}R)g_{\mu\nu,\alpha}$$
$$=\frac{\partial}{\partial x^{\tau}}\{\sqrt{-g}[L_{(G)}\delta^{\tau}_{\alpha}-(\frac{\partial L_{(G)}}{\partial g_{\mu\nu,\tau}}-\partial_{\sigma}\frac{\partial L_{(G)}}{\partial g_{\mu\nu,\tau\sigma}})g_{\mu\nu,\alpha}-\frac{\partial L_{(G)}}{\partial g_{\mu\nu,\tau\sigma}}g_{\mu\nu,\sigma\alpha}]\}$$
(6)

Using Eq. (1) and Eq. (5) and letting

$$t^{\tau}_{(G)\alpha} \stackrel{def}{=} L_{(G)}\delta^{\tau}_{\alpha}-(\frac{\partial L_{(G)}}{\partial g_{\mu\nu,\tau}}-\partial_{\sigma}\frac{\partial L_{(G)}}{\partial g_{\mu\nu,\tau\sigma}})g_{\mu\nu,\alpha}-\frac{\partial L_{(G)}}{\partial g_{\mu\nu,\tau\sigma}}g_{\mu\nu,\sigma\alpha},$$

the following identities can be derive from Eq. (6):

$$\frac{\partial}{\partial x^{\mu}}(\sqrt{-g}\,T^{\mu}_{(M)\nu}+\sqrt{-g}\,t^{\mu}_{(G)\nu})=0 \qquad (7)$$

Eq(7) can be considered as one kind of conservation laws for gravitational system including matter field and gravitational field, $t^{\mu}_{(G)\nu}$ is used to describe the energy-momentum for gravitational field, but it is not tensor.

The symmetries cited in the above statements, such as the action integrals $I_M$ and $I_G$, are all invariant under the local homogeneous Lorentz group and the local space-time translation group, $I_M$ is invariant also under internal symmetry, the three-dimensional space is maximally symmetric subspace of the whole of space-time, are all of the generally recognized symmetries of cosmos. These symmetries have some decisive effects on the evolution of cosmos.

In this paper we shall preserve these generally recognized symmetries of cosmos but revise the Lagrangian density of pure gravitational field and reconsider the conservation laws for gravitational system, so the statements (1) and (3) need to be changed while statement (2) is kept unchanged. Then, some theoretical foundations of the current cosmology are extended or revised and a new theory of cosmology is established. The differences between this new theory and the prevalent big bang cosmology are explored and discussed in this paper.

**2. The Modified Einstein's Field Equations**

We shall change the statement (1) to the statement (1'):

**Statement (1')**. The gravitational field variable is still $g_{\mu\nu}(x)$, the Lagrangian density of pure gravitational field is modified to be $\sqrt{-g}L_G(x)=-\frac{\sqrt{-g}}{16\pi G}[R(x)+2\lambda+2D(t)]$, where $\lambda$ is the cosmological constant, $D(t)$ is a scalar and is the function of time only in order to conform with the cosmological principle; from this $\sqrt{-g}L_G(x)$ the modified Einstein's field equations

$$R^{\mu\nu} - \frac{1}{2}g^{\mu\nu}R - \lambda g^{\mu\nu} - D^{\mu\nu} = -8\pi G T^{\mu\nu}_{(M)} \qquad (8)$$

are derived, where $D^{\mu\nu} \stackrel{def}{=} Dg^{\mu\nu}$ is a certain correction field which modifies Einstein's field equations; we also define: $T^{\mu\nu}_{(M)} \stackrel{def}{=} \frac{2}{\sqrt{-g}} \frac{\delta(\sqrt{-g}L_M)}{\delta g_{\mu\nu}}$, where $\sqrt{-g}L_M(x)$ is the Lagrangian density of matter field and it includes the interaction between matter field and gravitational field, so the energy-momentum tensor for matter field $T^{\mu\nu}_{(M)}$ includes the interaction energy between matter field and gravitational field. As a special case of Ref [17] it can be proved with the similar method that the action integral for matter field

$$I_M = \int \sqrt{-g} L_M d^4 x \qquad$$ and the action integral for gravitational field

$$I_G = \int \sqrt{-g} L_G d^4 x \qquad$$ are all invariant under the local nonhomogeneous Lorentz group. In

addition, $I_G$ is invariant only under space-time symmetry, but $I_M$ is invariant under both space-time symmetry and internal symmetry, so the gravitational field can be acted only by gravitational force and can not be acted by other forces, but the matter field can be acted by both gravitational force and other forces [6].

The steady state cosmology modifies firstly the Einstein's field equations by adding a correction tensor $D_{\mu\nu}$ [1], but the physical meaning and the method of introduction for $D^{\mu\nu}$ in this paper are different from the steady state cosmology. Below we shall show that the two terms $\lambda g^{\mu\nu}$ and $D^{\mu\nu}$ in our modified Einstein's field equations Eq. (8) would be well interpreted as 'dark energy' and a main part of 'dark matter' respectively. Due to these interpretations we should assume $\lambda > 0$ and $D(t) > 0$.

It must be emphasized that $\lambda g^{\mu\nu}$ and $D^{\mu\nu}$ in Eq. (8) are similar to $(R^{\mu\nu} - \frac{1}{2}g^{\mu\nu}R)$ in essence, they all are derived from the Lagrangian density for pure gravitational field $L_G$ and therefore they are quantities used to represent the pure gravitational field. It must be stressed that $I_G = \int \sqrt{-g} L_G d^4 x$ is invariant only under space-time symmetry, but $I_M = \int \sqrt{-g} L_M d^4 x$ is invariant under both space-time symmetry and internal symmetry; so the gravitational field can be acted only by gravitational force and can not be acted by other forces, but the matter field can be acted by both gravitational force and other forces. By virtue of $\lambda g^{\mu\nu}$ and $D^{\mu\nu}$ being quantities used to represent the pure gravitational field, these two parts of gravitational field can not interact with other forces including electromagnetic force, so they must be 'dark'; hence it is natural to interpret them as 'dark energy' and 'dark matter'.

Using the same method described in chapter 15 of reference [1], from Eqs. (2) (3) and (8) we can derive the following two equations:

$$(\frac{da}{dt})^2 + k = \frac{8\pi G}{3}(\rho_M + \frac{\lambda}{8\pi G} + \frac{D}{8\pi G})a^2 \qquad (9)$$

$$\frac{d^2 a}{dt^2} = -\frac{4\pi G}{3}(\rho_M + 3p_M - \frac{\lambda}{4\pi G} - \frac{D}{4\pi G})a \qquad (10)$$

where $\rho_M$ is the mean energy density of matter, and $p_M$ is the mean pressure of matter; the cosmological principle demands that $a(t), \rho_M(t), p_M(t), D(t)$ all depend on the cosmic standard time only [1]. We shall show in section 4 that at present time $(\rho_M + 3p_M - \frac{\lambda}{4\pi G} - \frac{D}{4\pi G}) < 0$, hence $\frac{d^2 a}{dt^2} > 0$, *i.e.* the universe is expanding acceleratively.

The four quantities $H(t), q(t), \rho_c, \Omega_{(M)}$ are used frequently in cosmology, their definitions are: $H(t) \overset{def}{=} \frac{da(t)/dt}{a(t)}; \quad q(t) \overset{def}{=} -\frac{d^2 a(t)}{dt^2} \frac{a(t)}{(\frac{da(t)}{dt})^2};$

when $t_0$ is the present time, the parameters $H_0 = H(t_0)$, $q_0 = q(t_0)$ are known as Hubble's constant and the deacceleration parameter respectively;

$$\rho_c \stackrel{def}{=} \frac{3H_0^2}{8\pi G} \ ; \qquad \Omega_{(M)} \stackrel{def}{=} \frac{\rho_M(t_0)}{\rho_c}$$

are called critical density and density parameter respectively. From Eqs.(9) and (10) and using these parameters the following relations can be obtained:

$$\frac{k}{H_0^2 a^2(t_0)} = (2q_0 - 1) + (\lambda + D(t_0))/H_0^2 \tag{11}$$

$$\Omega_{(M)} + \frac{1}{8\pi G \rho_c}(\lambda + D(t_0)) = 1 \tag{12}$$

To derive Eq. (11) we have used the fact that the matter energy density of the present university is dominated by nonrelativistic matter, so $p_M(t_0) \ll \rho_M(t_0)$ and $p_M(t_0)$ can be neglected. Eq. (11) implies that when

$$2q_0 = 1 - (\lambda + D(t_0))/H_0^2 \tag{13}$$

then $k = 0$. It has been determined $k=0$ from astronomical observations [7]; so Eq. (13) must be satisfied. If the two terms $\lambda g^{\mu\nu}$ and $D^{\mu\nu}$ in Eq. (8) do not exist, i.e. $\lambda = 0, D(t) = 0$, Eq. (13) becomes $2q_0 = 1$. If we define $\rho_\lambda \stackrel{def}{=} \frac{\lambda}{8\pi G}$, $\Omega_{(\lambda)} \stackrel{def}{=} \frac{\rho_\lambda}{\rho_c}$; $\rho_D \stackrel{def}{=} \frac{D(t)}{8\pi G}$,

$\Omega_{(D)} \stackrel{def}{=} \frac{\rho_D(t_0)}{\rho_c}$; then Eq. (12) becomes

$$\Omega_{(M)} + \Omega_{(\lambda)} + \Omega_{(D)} = 1 \tag{14}$$

Eq. (14) means that although $\lambda g^{\mu\nu}$ and $D^{\mu\nu}$ are two quantities which represent the pure gravitational field and are not two quantities which represent matter field in essence, but they have the property that they could be looked as if they are two parts of energy-momentum tensor of the matter fields. In order to explain this specific property we transform Esq. (8) into

$$R^{\mu\nu} - \frac{1}{2} g^{\mu\nu} R = -8\pi G T^{\mu\nu}_{mod} \tag{15}$$

where $T^{\mu\nu}_{mod}$ might be called modified energy-momentum tensor;

$$T^{\mu\nu}_{mod} \equiv T^{\mu\nu}_{(M)} - \frac{\lambda}{8\pi G} g^{\mu\nu} - \frac{D^{\mu\nu}}{8\pi G} \tag{16}$$

$T^{\mu\nu}_{mod}$ could also be written as the perfect-fluid form:

$$T^{\mu\nu}_{mod} = (\rho_{mod} + p_{mod}) u^\mu u^\nu + p_{mod} g^{\mu\nu} \tag{17}$$

comparing Eq. (16) with Eq. (17) we get

$$\rho_{mod} = \rho_M + \rho_\lambda + \rho_D; \quad \rho_\lambda = \frac{\lambda}{8\pi G}, \quad \rho_D = \frac{D}{8\pi G} \tag{18}$$

$$p_{mod} = p_M + p_\lambda + p_D; \quad p_\lambda = -\frac{\lambda}{8\pi G}, \quad p_D = -\frac{D}{8\pi G} \tag{19}$$

The relations for $\rho_\lambda$ and $\rho_D$ conform to the definitions of $\rho_\lambda$ and $\rho_D$ included in Eq. (14).

## 3. Lorentz and Levi-Civita's Conservation Laws of Energy-Momentum Tensor for Gravitational System

We shall change the statement (3) into the statement (3'):
**Statement (3')**. The contracted Bianchi identities [1]

$$(R^{\mu\nu} - \frac{1}{2} g^{\mu\nu} R)_{;\mu} = 0$$

hold under all conditions; thus $(R^{\mu\nu} - \frac{1}{2}g^{\mu\nu}R - \lambda g^{\mu\nu})_{;\mu} = 0$ and from Eq. (8) we must have

$$(T^{\mu\nu}_{(M)} - \frac{D^{\mu\nu}}{8\pi G})_{;\mu} = 0 \ , \tag{20}$$

but owing to $\frac{d}{dt}D(t) \neq 0$, there is no identities $D^{\mu\nu}_{;\mu} = 0$, hence the identities $(T^{\mu\nu}_{(M)})_{;\mu} = 0$ do not exist either, *i.e.*

$$T^{\mu}_{(M)\nu;\mu} = \frac{1}{\sqrt{-g}}\frac{\partial(\sqrt{-g}T^{\mu}_{(M)\nu})}{\partial x^{\mu}} - \frac{1}{2}\frac{\partial g_{\mu\rho}}{\partial x^{\nu}}T^{\mu\rho}_{(M)} \neq 0 \ ; \tag{21}$$

from Eqs. (20, 21) we have

$$\frac{1}{\sqrt{-g}}\frac{\partial(\sqrt{-g}T^{\mu}_{(M)\nu})}{\partial x^{\mu}} - \frac{1}{2}\frac{\partial g_{\mu\rho}}{\partial x^{\nu}}T^{\mu\rho}_{(M)} - \frac{1}{8\pi G}D_{,\nu} = 0 \ . \tag{22}$$

On the other hand, because $I_G = \int \sqrt{-g} L_G d^4x$ is invariant under the local space-time translation group, there are the other identities

$$\frac{\sqrt{-g}}{16\pi G}(R^{\mu\nu} - \frac{1}{2}g^{\mu\nu}R - \lambda g^{\mu\nu} - D^{\mu\nu})g_{\mu\nu,\alpha}$$

$$= \frac{\partial}{\partial x^{\tau}}\{\sqrt{-g}[L_{(G)}\delta^{\tau}_{\alpha} - (\frac{\partial L_{(G)}}{\partial g_{\mu\nu,\tau}} - \partial_{\sigma}\frac{\partial L_{(G)}}{\partial g_{\mu\nu,\tau\sigma}})g_{\mu\nu,\alpha} - \frac{\partial L_{(G)}}{\partial g_{\mu\nu,\tau\sigma}}g_{\mu\nu,\sigma\alpha}]\}$$

$$+ \frac{1}{8\pi G}\sqrt{-g}D_{,\alpha}$$

$$\tag{23}$$

Eq. (23) can be proved with the similar method of Ref. [10]. By using Eq. (8) and Eq. (22) and defining

$$t^{\tau}_{(G)\alpha} \stackrel{def}{=} L_{(G)}\delta^{\tau}_{\alpha} - (\frac{\partial L_{(G)}}{\partial g_{\mu\nu,\tau}} - \partial_{\sigma}\frac{\partial L_{(G)}}{\partial g_{\mu\nu,\tau\sigma}})g_{\mu\nu,\alpha} - \frac{\partial L_{(G)}}{\partial g_{\mu\nu,\tau\sigma}}g_{\mu\nu,\sigma\alpha} \ ,$$

from Eq.(23) the identities Eq.(7)

$$\frac{\partial}{\partial x^\mu}(\sqrt{-g}\,T^\mu_{(M)\nu}+\sqrt{-g}\,t^\mu_{(G)\nu})=0$$

can also be derived.

Einstein maintained that [8] $t^\mu_{(G)\nu}$ should be used to describe the energy-momentum for gravitational field and Eq.(7) should be regarded as the conservation laws for gravitational system including matter field and gravitational field; so we shall call Eq.(7) Einstein's conservation laws.

But the quantity $t^\mu_{(G)\nu}$ has two serious defects: first, $t^\mu_{(G)\nu}$ is not tensor, it lacks the invariant character required by the principles of general relativity [8,9]; secondly, many quantities $t^{*\mu}_{(G)\nu}$ could be used instead of $t^\mu_{(G)\nu}$, if

$$\sqrt{-g}\,t^{*\mu}_{(G)\nu}=\sqrt{-g}\,t^\mu_{(G)\nu}+\partial_\sigma(\sqrt{-g}\,u^{\mu\sigma}_\nu)\;,\quad u^{\mu\sigma}_\nu$$ is any quantity which satisfies

$u^{\mu\sigma}_\nu=-u^{\sigma\mu}_\nu$, then $t^{*\mu}_{(G)\nu}$ also satisfies Eq.(7). On the other hand it can be proved in the appendix that there exist the relations [10]

$$\sqrt{-g}\,t^\mu_{(G)\nu}=2\frac{\delta(\sqrt{-g}\,L_G)}{\delta g_{\mu\lambda}}g_{\lambda\nu}-\partial_\sigma(\sqrt{-g}\,V^{\mu\sigma}_{(G)\nu})\;,\text{ and}$$

$$\partial_\sigma(\sqrt{-g}\,V^{\mu\sigma}_\nu)=-\partial_\sigma(\sqrt{-g}\,V^{\sigma\mu}_\nu)\;,\text{ then}$$

$$\sqrt{-g}\,t^{*\mu}_{(G)\nu}=2\frac{(\delta\sqrt{-g}\,L_G)}{\delta g_{\mu\lambda}}g_{\lambda\nu}-\partial_\sigma(\sqrt{-g}\,V^{\mu\sigma}_{(G)\nu})+\partial_\sigma(\sqrt{-g}\,u^{\mu\sigma}_\nu)\;;$$

therefore if we choose a quantity $u^{\mu\sigma}_\nu$ so that $\partial_\sigma(\sqrt{-g}\,V^{\mu\sigma}_{(G)\nu})+\partial_\sigma(\sqrt{-g}\,u^{\mu\sigma}_\nu)=0$,

consequently $\sqrt{-g}\,t^{*\mu}_{(G)\nu}=2\dfrac{(\delta\sqrt{-g}\,L_G)}{\delta g_{\mu\lambda}}g_{\lambda\nu}$, this is a special case of $\sqrt{-g}\,t^{*\mu}_{(G)\nu}$.

$\dfrac{2}{\sqrt{-g}}\dfrac{(\delta\sqrt{-g}L_G)}{\delta g_{\mu\lambda}}g_{\lambda\nu}$ is a tensor, it does not have the two serious defects cited above. To follow the definition of the energy-momentum tensor for the matter field $T^{\mu\nu}_{(M)}\overset{def}{=}\dfrac{2}{\sqrt{-g}}\dfrac{\delta(\sqrt{-g}L_M)}{\delta g_{\mu\nu}}$, Lorentz and Levi-Civita had defined [8] the energy-momentum tensor for the gravitational field by $T^{\mu\nu}_{(G)}\overset{def}{=}\dfrac{2}{\sqrt{-g}}\dfrac{\delta(\sqrt{-g}L_G)}{\delta g_{\mu\nu}}$, thus they obtained the conservation laws of energy-momentum tensor for gravitational system including matter fields and gravitational fields as:

$$\dfrac{\partial}{\partial x^\mu}(\sqrt{-g}T_{(M)\mu\nu}+\sqrt{-g}T_{(G)\mu\nu})=0 \qquad (24)$$

and $$T_{(M)\mu\nu}+T_{(G)\mu\nu}=0 \qquad (25)$$

Eq. (25) originates from the field equations and the definition of the energy-momentum tensor for gravitational field. We shall call Eq. (24) and Eq. (25) Lorentz and Levi-Civita's conservation laws.

Einstein did not agree with Lorentz and Levi-Civita's conservation laws for the sole reason that Eq. (25) "does not exclude the possibility that a material system disappears completely, leaving no trace of its existence."[8], because Einstein believed that the relation expressed by Eq. (25) should make the energy-momentum of a material system, being $T_{(M)\mu\nu}\neq 0$ in the initial state, to $T_{(M)\mu\nu}\to 0$ spontaneously. We shall show that this view is not correct. According to statistical mechanics the entropy $S$ of a macroscopic system must obey the Boltzmann's relation $S=k\ln N$, where $N$ is the number of microscopic states. For a macroscopic system, there must be $N\gg 1$ always, thus $S>0$ usually. If a gravitational system (including matter and gravitational field) could disappear completely and spontaneously, then in the disappearing process $N$ will decrease to $N=1$ gradually; here we look upon the complete disappearance as a special state. Because there is no difference in the meaning between macroscopic and microscopic state for the complete disappearance, so $N=1$. Therefore in the complete disappearing process of this gravitational system its entropy should decrease to $S=0$ from $S>0$; this is contrary to the theorem of entropy increase; hence a gravitational system can not disappear completely and spontaneously.

In the last few years I have thoroughly studied Lorentz and Levi-Civita's conservation laws and found that these conservation laws not only are rational and perfect but also have abundant physical contents [10-13]. A number of new specific properties of gravitational fields or gravitational waves can be deduced and can be tested via experiments or observations [13]. The Lorentz and Levi-Civita's conservation laws will be used as one of important theoretical foundations to establish a new cosmology.

### 4. A New Theory of Cosmology

Using the statements [1'][2'][3'] as the primary theoretical foundations we shall establish a new theory of cosmology which is different from the prevalent big bang cosmology in the following aspects:

(1). In this new theory of cosmology, the Lagrangian density of pure gravitational field is modified to be $\sqrt{-g}L_G(x) = -\frac{\sqrt{-g}}{16\pi G}[R(x) + 2\lambda + 2D(t)]$, so it adopts the modified Einstein equations (8)

$$R^{\mu\nu} - \frac{1}{2}g^{\mu\nu}R - \lambda g^{\mu\nu} - D^{\mu\nu} = -8\pi G T^{\mu\nu}_{(M)}$$ as the equations of gravitational field for the universe; but in the big bang cosmology, $R^{\mu\nu} - \frac{1}{2}g^{\mu\nu}R = -8\pi G T^{\mu\nu}_{(M)}$ or

$R^{\mu\nu} - \frac{1}{2}g^{\mu\nu}R - \lambda g^{\mu\nu} = -8\pi G T^{\mu\nu}_{(M)}$ is adopted as the equations of gravitational field for the universe.

(2). This new theory of cosmology affirms that $\lambda g^{\mu\nu}$ and $D^{\mu\nu}$ are two quantities used to represent the pure gravitational field and not used to represent matter field in essence, $\lambda g^{\mu\nu}$ and $D^{\mu\nu}$ are interpreted as dark energy and a main part of dark matter respectively; but in the big bang cosmology, although $\lambda g^{\mu\nu}$ is interpreted as dark energy also, yet it is always considered to be a part of matter field; especially there is no quantity $D^{\mu\nu}$ in the big bang cosmology, the dark matter is looked entirely as some kinds material matter.

(3). In this new theory of cosmology, the Lorentz and Levi-Civita's conservation laws

$$\frac{\partial}{\partial x^\mu}(\sqrt{-g}T_{(M)\mu\nu} + \sqrt{-g}T_{(G)\mu\nu}) = 0$$

$$T_{(M)\mu\nu} + T_{(G)\mu\nu} = 0$$

are used to study the changes taking place in a gravitational system; from these laws we shall show in the following that the energy-momentum of matter field might 'create' from gravitational field; but in the big bang cosmology, Einstein's conservation laws

$$\frac{\partial}{\partial x^\mu}(\sqrt{-g}\,T^\mu_{(M)\nu}+\sqrt{-g}\,t^\mu_{(G)\nu})=0$$ are used to study the changes taking place in a

gravitational system; besides, the big bang cosmology supposes that the total energy of matter fields (including the inflation field) has existed from the big bang, it does not study the origin of the matter field's energy.

Below we shall present detailed studies on these differences and give a new interpretation on the evolution of cosmos.

### 4.1 Dark matter and dark energy

The new theory of cosmology advanced in this paper holds the view that there are two kinds of dark matter: one should be the field of $D^{\mu\nu}$ which energy density is $\rho_D$, and the other might be some material matter [7], such as the neutrino, a weakly interacting massive particle (WIMP) and the massive compact halo objects (MACHOs, including low-luminosity stars and black holes), *etc.;* their energy density are some parts of $\rho_M$. The conclusions from CMB data tell us that [7] the Universe is made up as follows: 73% dark energy, 23% dark matter and 4% ordinary (baryonic) matter. According the above view point we would have:

$$\rho_\lambda(t_0)/\rho_c = 73\%\quad,\quad \rho_M(t_0)/\rho_c > 4\%\quad,\quad \rho_D(t_0)/\rho_c < 23\%\quad,\quad \text{and}$$

$$\rho_M(t_0)/\rho_c + \rho_D(t_0)/\rho_c = 27\%.$$ In this new theory of cosmology, we can distinguish between $\rho_D$ and $\rho_M$; since $\rho_D$ can be only acted by gravitational force and can not be acted by other forces, but $\rho_M$ can be acted by both gravitational force and other forces, hence it could be possible to distinguish the two kinds of dark matter. These possibilities might be tested by experiments and observations in future.

It must be pointed out, for the whole cosmos $\rho_\lambda, \rho_D, \rho_M$ are all less than the critical density [1]

$$\rho_c = \frac{3[H(t_0)]^2}{8\pi G} = 1.9 h^2 \times 10^{-29}\,g/cm^3\;;$$ but for a macroscopic gravitational system,

$\rho_M \gg \rho_C$, however $\rho_\lambda, \rho_D$ still less than $\rho_C$, then from Eq. (18) $\rho_{mod} \approx \rho_M$, therefore Eq. (8)

$$R^{\mu\nu} - \frac{1}{2}g^{\mu\nu}R - \lambda g^{\mu\nu} - D^{\mu\nu} = -8\pi G T^{\mu\nu}_{(M)} \text{ degenerate to}$$

$$R_{\mu\nu} - \frac{1}{2}g_{\mu\nu}R = -8\pi G T_{(M)\mu\nu} .$$

The Eq.(10) can be rewritten as $\frac{d^2 a}{dt^2} = -\frac{4\pi G}{3}(\rho_M + 3p_M - 2\rho_\lambda - 2\rho_D)a$;

owing to $p_M(t_0) \ll \rho_M(t_0)$ and utilizing the above CMB data, it is evidently

$(\rho_M + 3p_M - 2\rho_\lambda - 2\rho_D) < 0$, therefore $\frac{d^2 a}{dt^2} > 0$, *i.e.* the universe is accelerating in its expansion.

It had been suggested by some scholars that the energy density $\rho_\lambda$ of field $-\frac{\lambda}{8\pi G}g_{\mu\nu}$ perhaps might be equal to the vacuum energy density of matter field [14]; but their views all run into some difficulties and conflicting issues. In this new theory of cosmology, $\rho_\lambda$ is a part of gravitational field's energy density, but the vacuum energy density of matter field is a part of matter field's energy density $\rho_M$ which belongs to $T^{\mu\nu}_{(M)}$; they might be different in essence, and there is no relation between $\rho_\lambda$ and the vacuum energy density of matter field. We have indicated above that $\rho_\lambda$ can be only acted by gravitational force and can not be acted by other forces, but $\rho_M$ can be acted by both gravitational force and other forces, so it could be possible to distinguish $\rho_\lambda$ from the vacuum energy density of matter field by future experiments and observations.

**4.2 Some deductions from the Lorentz and Levi-Civita's conservation laws, the origin of matter field's energy.**

From Eq. (25) we get $\triangle T_{(M)\mu\nu} = -\triangle T_{(G)\mu\nu}$ immediately, this relation means that for an isolated gravitational system if the energy-momentum of matter field increases, then the energy-momentum of gravitational field should decrease, *i.e.* the energy-momentum of gravitational field might transform into the energy-momentum of matter field. This possibility might occur in reality, since the number of microscopic states both for matter field and gravitational field should all increase in this process so that the entropy of the system increases. It is worth pointing out that in the above process the absolute value of gravitational field energy is increasing, thus the number of microscopic states for gravitational field should increase also. This possibility could be used to explain the origin of matter field's energy and this explanation is one important characteristic of our new theory of cosmology. Before discussing the origin of matter field's energy, we shall deduce some relations from the Lorentz and Levi-Civita's conservation laws first.

Comparing Eq.( 8) with Eq. (25), we get

$$T_{(G)}^{\mu\nu} = \frac{1}{8\pi G}(R^{\mu\nu} - \frac{1}{2}g^{\mu\nu}R - \lambda g^{\mu\nu} - D^{\mu\nu}) \tag{26}$$

This equality means $T_{(G)}^{\mu\nu}$ can be divided into three parts:

$$T_{(G)}^{\mu\nu} = {}^R T_{(G)}\mu\nu + {}^\lambda T_{(G)}\mu\nu + {}^D T_{(G)}\mu\nu \tag{27}$$

where ${}^R T_{(G)}\mu\nu = \frac{1}{8\pi G}(R^{\mu\nu} - \frac{1}{2}g^{\mu\nu}R)$ is the part of gravitational field's energy-momentum due to space-time curvature; ${}^\lambda T_{(G)}\mu\nu = -\frac{\lambda g^{\mu\nu}}{8\pi G}$ is the part of gravitational field's energy-momentum due to cosmological constant; ${}^D T_{(G)}\mu\nu = -\frac{D^{\mu\nu}}{8\pi G}$ is the part of gravitational field's energy-momentum due to the correction field $D^{\mu\nu}$.

From Eqs. (8, 24-27) we obtain

$$\overset{R}{T}_{(G)}\mu\nu + \overset{\lambda}{T}_{(G)}\mu\nu + \overset{D}{T}_{(G)}\mu\nu + T_{(M)}{}^{\mu\nu} = 0 \tag{28}$$

$$\frac{\partial}{\partial x^{\mu}}(\overset{R}{T}_{(G)}\mu\nu + \overset{\lambda}{T}_{(G)}\mu\nu + \overset{D}{T}_{(G)}\mu\nu + T_{(M)}{}^{\mu\nu}) = 0 \tag{29}$$

Let $\mu = \nu = 0$ in Eqs. (28, 29) then we have

$$\rho_R + \rho_\lambda + \rho_D + \rho_M = 0 \tag{30}$$

$$\frac{d}{dt}(\rho_R + \rho_\lambda + \rho_D + \rho_M) = 0 \tag{31}$$

$\rho_G = \rho_R + \rho_\lambda + \rho_D$ is the total energy density of the pure gravitational field, $\rho_R, \rho_\lambda$ or $\rho_D$ is the energy density relating to $\overset{R}{T}_{(G)}\mu\nu$, $\overset{\lambda}{T}_{(G)}\mu\nu$ or $\overset{D}{T}_{(G)}\mu\nu$ respectively.

This new theory of cosmology assumes that $\rho_\lambda \geq 0$, $\rho_D \geq 0$ and $\rho_M \geq 0$ always, hence $\rho_R \leq 0$ and $\rho_G = \rho_R + \rho_\lambda + \rho_D \leq 0$ always. On the other hand, from Eqs(15-20) and using the same methods of Ref.[1], we can get the relations:

$$\frac{d(\rho_M + \rho_D)}{dt} + 3\frac{\frac{da}{dt}}{a}(\rho_M + \rho_D + p_M + p_D) = 0 \tag{32}$$

Eq. (32) is equivalent to Eq. (31), there exists the relation:

$$\frac{d(\rho_R + \rho_\lambda)}{dt} = 3\frac{\frac{da}{dt}}{a}(\rho_M + \rho_D + p_M + p_D), \text{ i.e. } \frac{d\rho_R}{dt} = 3\frac{\frac{da}{dt}}{a}(\rho_M + p_M).$$

Since Eqs (18, 19) tell us $\rho_D + p_D = 0$, besides, $3\frac{da/dt}{a} \cong \frac{\Delta V}{V \Delta t}$, so the Eq.(32) can be rewritten as

$$\frac{\Delta \rho_D}{\Delta t} \cong -\frac{\Delta(\rho_M V)}{V \Delta t} - p_M \frac{\Delta V/\Delta t}{V}, \qquad (33)$$

where V is any volume in the space, $\frac{\Delta \rho_D}{\Delta t}$ represents the rate of energy density change for the field $D^{\mu\nu}$, $\frac{\Delta(\rho_M V)}{V \Delta t} = \frac{\Delta \rho_M}{\Delta t} + \frac{\rho_M \Delta V}{V \Delta t}$ represents the total energy change per unit volume per unit time for the matter field, $p_M \frac{\Delta V/\Delta t}{V}$ represents the work done per unit volume per unit time by the matter field. For matter field, there are $\rho_M \geq 0$ and $p_M \geq 0$ usually; if $\Delta V > 0$, $\Delta \rho_M > 0$ or $\Delta \rho_M < 0$ but $\Delta \rho_M + \frac{\rho_M \Delta V}{V} > 0$, then $\frac{\Delta(\rho_M V)}{V \Delta t} > 0$, *i.e.* the energy of the matter field increases, and $p_M \frac{\Delta V}{V \Delta t} > 0$, *i.e.* the work done by the matter field is positive. Hence from Eq. (33) $\frac{\Delta \rho_D}{\Delta t} < 0$, this relation means that some energy of field $D^{\mu\nu}$ has transformed into the energy of matter field. This tells us that the increase of matter field energy stems the decrease of gravitational field energy.

If we assume that at initial time $t=0$, $\rho_M = 0$, $p_M = 0$ everywhere, *i.e.*

$T^{\mu\nu}{}_{(M)}(0) = 0$ everywhere (Since $T^{\mu\nu}{}_{(M)} = p_M g^{\mu\nu} + (\rho_M + p_M)U^\mu U^\nu$); then according to the above analysis, the energy of matter field would be transformed from the gravitational field continuously, this means that the energy of matter field might originate from the gravitational field.

The state $T^{\mu\nu}{}_{(M)}(0) = 0$ is the lowest state of energy-momentum for the matter field in the universe. It must be emphasized that this state is not equal to the other lower energy state, *i.e.* the so called 'vacuum' state of quantum matter field; since at the 'vacuum' state, $\rho_M > 0$.

On the other hand, it must be indicated that the energy creation of matter field does not mean the matter field creation, thus if at $t=0$, $\rho_M(0) = 0$, at $t>0$, $\rho_M(t) > 0$, it means only that the state of matter field is change from the lowest state to a higher state, but the matter field exists all along from the beginning of the energy change. It must be indicated also that our new theory of cosmology set forth in this paper has no beginning state with $T^{\mu\nu}{}_{(M)}(0) = 0$ everywhere in the space, *i.e.* the state $t=0$, $\rho_M = 0$, $p_M = 0$ everywhere does not exist. Why is there not this beginning state ? This is due to the quantum fluctuations, at any time there must always be energy-momentum transformations between gravitational field and matter field, so the beginning state $t=0$, $\rho_M = 0$, $p_M = 0$ is not possible. Moreover we could assume that the quantum fluctuations might have an infinite existence time.

**4.3 The big bang may never have occurred**

The standard cosmology (SBBC) has a beginning state called big bang, and it is assumed that the total energy of matter fields (including the inflation field) has existed since the big bang. At the big bang, *i.e.* at $t=0$, it is generally thought that $\rho_M \to \infty$ and the temperature $T \to \infty$. Moreover, SBBC does not study the origin of the matter field's energy. As we have shown in the above discussions, the energy of matter field might be transformed from the gravitational field continuously, this means that at $t=0$, the state $\rho_M \to \infty$ does not exist, therefore the big bang might never have occurred.

In addition, in SBBC it is always thought that $a(t) = 0$ at the big bang and the time elapse of the universe expansion from the big bang to the present is finite; but in our new theory of cosmology, the time elapse of the universe expansion might have beem infinite. This is because in the new theory we can always suppose that $a(t) \geq 0$, $\frac{d}{dt}a(t) \geq 0$ and $\frac{d^2}{dt^2}a(t) \geq 0$, hense by Law of the mean we have $a(t_0) - a(t_I) = (t_0 - t_I)[\frac{d}{dt}a(t)]_{t=t_m}$, where $t_I < t_m < t_0$; let $t_0$ be the present time and suppose $a(t_I) = 0$, subsequently we get

$$t_0 - t_I = a(t_0) \Big/ [\frac{d}{dt}a(t)]_{t=t_m} \text{ ; thus if } [\frac{d}{dt}a(t)]_{t=t_m} \to 0,$$

then $t_0 - t_I \to \infty$, *i.e.* it is possible that the time elapse of the universe expansion might have beem infinite.

Owing to the reasons cited above the new theory of cosmology set forth in this paper holds that the universe is without a beginning and without an end, and space expands continuously.

How does the energy-momentum transform from the gravitational field into the matter field? This problem relates to quantum theory of gravitational field. As a complete and consistent quantum theory of gravitational field has not yet been constructed yet, we can not answer this problem clearly and completely; however, we could put forward the following assumptions which will be proved , or refuted, or revised by future experiments and observations:

(1). The energy of gravitational field might transform into the energy of some elementary particles (including the thermal energy of elementary particles); but these transformed energy can not lead to the state $\rho_M \to \infty$ and the temperature can not reach $T \to \infty$.

(2). In the past, when some conditions were satisfied; some eras, which were similar to the eras of the early universe in SBBC [1], might emerge from the quantum fluctuation. But in our new theory of cosmology, the period of every era might be very long and the cosmic changes taking place in the matter field might be very slow. As an example We shall use Eq.(33) to show that the cosmic change taken place in the matter field for the radiation-dominated era:

Rewriting Eq.(33) $\frac{\Delta \rho_D}{\Delta t} \cong -\frac{\Delta(\rho_M V)}{V \Delta t} - p_M \frac{\Delta V / \Delta t}{V}$ as

$$\frac{d\rho_D}{dt} = -\frac{d(\rho_M V)}{V dt} - p_M \frac{dV/dt}{V}. \tag{33'}$$

For the radiation-dominated era, $p_M(t) = \frac{1}{3}\rho_M(t)$, Eq.(33') become

$\frac{d\rho_M}{\rho_M} + 4\frac{da}{a} = -d\rho_D$. In SBBC, $\rho_D = 0$, we shall get $\rho_M a^4 = 1$; in the new theory of cosmology, if $dV/dt > 0$, $d\rho_M/dt < 0$ but $\frac{d(\rho_M V)}{V dt} > 0$, then $d\rho_D < 0$, we shall get $\rho_M a^4 > 1$. It is obvious that, when the universe expands, $\rho_M$ will decrease slower in the new theory of cosmology than in SBBC.

(3). Especially, there had been the change from the radiation-dominated era to the matter-dominated era which is similar with SBBC. At the radiation-dominated era, matter and radiation were presumably in thermal equilibrium; their temperature is higher than $4000\ ^oK$. When the temperature is below $4000\ ^oK$, the matter-dominated era commenced, and the radiation existed still and had been red-shifted owing to the expansion of the universe. It is widely believed that the microwave radiation background is just the left-over radiation [1]; so our new theory of cosmology can explain the microwave radiation background as well as SBBC.

In SBBC the observed abundances of light nuclei in the universe are explained as the result of nucleon-synthesis taking place in the early universe at a temperature of about $10^9\ ^oK$. In the new theory of cosmology, although the observed abundances of light nuclei in the universe can be explained with the same reason as SBBC, there is another explanation which had put forward by some cosmologists in the 1950's. They had studied the possibilities of that the light nuclei in the universe are formed from hydrogen nuclei in the interiors of stars [1]; but the cosmic abundance of helium is too large to be easily explained in terms of nucleon-synthesis in the interiors of stars at $10^{10}$ years estimated by SBBC. However the new theory of cosmology is without a beginning state, the helium nuclei in the universe might be synthesized in a longer time frame; therefore this problem does not exist. Which explanation is correct will be determined by future tests.

5. **Some Future Tests**

The validity of a theory must be tested by experiments and observations. We outline some future tests that could either confirm or disprove the proposed new theory of cosmology in the following:

**5.1 Testing the Lorentz and Levi-Civita's conservation laws**

Various experiments and observations using the specific properties of gravitational waves to test the Lorentz and Levi-Civita's conservation laws have been described in Ref. [13]. These conservation laws are the foundation of our new theory; their correctness means that the energy-momentum of the matter field might create from gravitational field. So that to confirm these conservation laws is to confirm indirectly the new theory of cosmology and to disprove SBBC, since SBBC does not permit the creation of matter field's energy-momentum from gravitational field.

**5.2 Investigating the essence of the dark matter**

We have explained in the above discussions that in the new theory of cosmology some parts of the 'dark matter' might be material matter, the other part of the dark matter should be the field of $D_{\mu\nu}$, which is a part of gravitational field. The gravitational field is different from the matter field, whether $\rho_D$ is material matter can be distinguished by interactions. To confirm $D_{\mu\nu}$ being a part of gravitational field is to confirm the new theory of cosmology and to disprove SBBC.

**5.3 Investigating the essence of the dark energy**

We have also indicated above that $-\lambda g_{\mu\nu}$ is a part of gravitational field, whether $\rho_\lambda$ is material matter can be distinguished by interactions. To confirm $-\lambda g_{\mu\nu}$ being a part of gravitational field is to confirm the new theory of cosmology and to disprove SBBC.

**5.4 Searching very old stars**

The new theory of cosmology is without a beginning and without an end; therefore very old stars might exist. To find very old stars is to confirm the new theory of cosmology and to disprove SBBC.

Finally I would like to comment about the purpose of this paper. As many new evidences of observations [7, 15, 16] have brought out some crucial weaknesses of SBBC. It is necessary to introduce new concepts and new theories; the goal of this paper is an attempt to satisfy this need. Perhaps the theories set forth in this paper are not yet perfect, but I believe that it might have significant consulting value.